# Modeling the Role of Secondary Electron Emission in Direct Current Magnetron Sputtering using Explicit Energy-Conserving Particle-in-Cell Methods

Daniel Main[a], Thomas G. Jenkins[a], Joseph G. Theis[b], Gregory R. Werner[b], John R. Cary[a,b], Eve Lanham[a,c], Seth A. Veitzer[a], Scott E. Kruger[a]

a. Tech-X Corporation (now part of Silvaco), 5621 Arapahoe Ave suite a, Boulder, CO 80303
b. Center for Integrated Plasma Studies, University of Colorado, Boulder, CO 80309-0390
c. Currently at Lam Research, 11355 SW Leveton Dr, Tualatin, OR 97062

**Abstract**

We present results from a fully kinetic particle-in-cell (PIC) simulation of direct current magnetron sputtering (dcMS) in a 2D cylindrically symmetric geometry. The particle-in-cell model assumes an electrostatic approximation and includes the Monte Carlo collision (MCC) method to model collisions between electrons and the neutral gas. A newly-implemented explicit energy-conserving PIC algorithm (EC-PIC) is also exercised by the model and results are compared with the standard momentum conserving PIC (MC-PIC) method. We use these simulation tools to examine how changes in secondary electron yield (SEY) and the external circuit impact the steady-state current, voltage, and plasma density of dcMS discharges. We show that in general, higher SEY and lower external resistance values lead to larger currents, smaller voltages, and larger plasma densities. We demonstrate that EC-PIC is superior to MC-PIC as the plasma current and density increase due to the improved numerical stability provided by EC-PIC.

**Introduction**

*Magnetron Sputtering*

Magnetron sputtering is a physical vapor deposition (PVD)[1] technique that has become an essential technology for thin film deposition in a wide range of industrial applications[2]. Such industrial uses include metallization in integrated circuits[3], coatings for wear resistance and corrosion protection[4], optical coatings[5], and photovoltaic solar cells[6]. Magnetron sputtering began as an industrial technique in the 1960s as a replacement for diode sputtering devices[2], which suffered from low deposition rates and high voltage source requirements. The original magnetron sputtering configuration is known today as direct current magnetron sputtering (dcMS) and operates by applying a DC voltage to a cathode. The (negative) cathode voltage is typically of order 300 – 700 V. A magnetic field created from magnets placed behind the cathode, together with a sheath electric field, serve to confine electrons near the cathode. Therefore, almost all the ionization in the device occurs within a region next to the cathode target. The plasma discharge region has a typical density of $10^{15}$-$10^{17}$ m$^{-3}$. The neutral pressure in most dcMS systems is 1-30 mTorr, which implies that the mean free path is larger than most devices. Without the magnetic field, electrons would travel to the anode with a very small probability of ionizing the background gas. The magnetic field inhibits this transport and increases the time an electron spends in the discharge region, thus increasing the number of ionizing collisions it can undergo before impacting a boundary. In the discharge, ions are accelerated toward a target (usually the cathode of the discharge), sputtering off cathode material to coat the desired substrate. Other important processes include ion-induced and electron-induced secondary electron emission from the cathode. These two physical processes are independent. In this paper, we investigate the role of ion-induced secondary electron emission on the evolution of dcMS discharges.

The numerical modeling of dcMS discharges can streamline the design of physical magnetron sputtering devices, enabling process conditions such as magnetic field strength and configuration, neutral pressure, and operating voltage to be optimized. Optimizing these parameters can improve the deposition rate and uniformity onto the substrate and increase the device efficiency, in addition to tuning other desired properties. A variety of numerical models (global 0D models [7], multi-fluid models [8], hybrid models [fluid ions, kinetic electrons] [9]), and fully kinetic models (further discussion provided later in this section) may provide useful design information, though not every model captures the discharge physics in detail. At low pressures (< ~ 30 mT), for instance, only fully kinetic models can accurately represent the non-Maxwellian ion distribution function [10]. Likewise, the evolution of the plasma-gas mixture at low pressure (which exhibits behavior in the "free molecular flow" regime) is better modeled with a Monte Carlo approach rather than with fluid rate equations [11].

In this paper we will discuss dcMS discharge modeling results from an explicit PIC-MCC kinetic plasma simulation model called VSim [12, 13]. Although VSim allows for a large variety of PIC algorithms and geometries, we used its explicit FDTD electrostatic PIC-MCC capabilities with 2D (R-Z) cylindrical geometry. The Monte Carlo collision algorithm [14] is used for plasma-neutral interactions. The focus of this work is twofold: (a) to assess the performance of VSim's explicit energy-conserving PIC algorithm (EC-PIC) relative to standard momentum-conserving PIC (MC-PIC) algorithms, and (b) to investigate how the ion-induced secondary electron yield at the cathode and external circuit affect the plasma potential, cathode current, and resulting sputtering yield.

Previous modeling work on dcMS discharges using the PIC-MCC method includes that of Hua-Yu & Zong-Xin [15], which focuses on cathode erosion in planar discharges; Pflug et al. [16] in which ion energy distribution functions impacting the cathode are explored as discharge power sources are varied (RF vs. DC); and Ekpe et al. [17], in which energy fluxes and neutral deposition rates on the substrate are compared as magnetic field strength is varied. Ryabinkin et al. [18] examined plasma properties in a dcMS system as a function of gas pressure, using under-resolved grid spacing in the radial direction mitigated by a Gaussian smoothing method. Other PIC-MCC magnetron studies include Kolev et al. [19], who explored the role of electron recapture at the cathode. More recently, Theis et al. [20] have also used VSim to reproduce voltage-pressure curves obtained from dcMS experiments [21].

*dcMS discharge modeling: Numerical Considerations*
To optimize an industrial magnetron sputtering system, it is necessary to quantify the uniformity and rate of the sputtering deposition onto the substrate for a given configuration. Numerically, the determination of these quantities is challenging due to the significant disparities in timescales and spatial scales associated with various physical processes in the discharge. Accurately determining the sputtering deposition rate and uniformity requires accurate modeling of the plasma and sheath near the cathode, since it is there that the sputtering events occur. The energy of individual sputtering events, arising from ions impacting the cathode, is highly sensitive to sheath parameters. Plasma and sheath dynamics on the shortest time-scales are governed primarily by electrons, and the kinetic modeling of electron evolution via conventional MC-PIC methods is restrictive, requiring both spatial resolution of the electron Debye length ($\lambda_{De}$) and time resolution of the electron plasma frequency to avoid numerical instability. Accordingly, the simulation must resolve timescales for processes both short (~10 ps, to resolve electron motion), intermediate (~10 μs, for ion transport effects to bring the plasma discharge to steady-state), and long (~10 ms, to bring the sputtered neutrals to the substrate). Analogous disparities in spatial scale must also be resolved, ranging from the electron Debye length ($\lambda_{De}$ ~ 0.02 – 0.5 mm) to the size of the chamber itself (~10 cm or greater).

To address the timescale difficulties, we have developed a two-step simulation workflow, in which we separately model (1) the steady-state plasma formation and (2) the ensuing sputtered neutral flux from the cathode. In the first step, we model the plasma, sheath, and sputtered neutral evolution until steady state is reached. We then begin a second simulation in which the sputtered neutral flux from the first simulation is used as a boundary condition. This second simulation does not include the plasma dynamics; it can therefore have much larger time steps and grid sizes and run correspondingly faster. Using this two-step workflow, we can model the sputtered flux onto the substrate. Furthermore, we can model the erosion rate of the cathode and estimate how long the cathode material can be used before it will need to be replaced. This paper will focus on step 1 of this workflow, while a planned companion paper will focus on step 2.

To address the difficulties posed by disparate spatial scales, a number of approaches can be used. In a typical dcMS setup, the electron Debye length ranges from ~ 0.02 – 0.5 mm; for a chamber of characteristic length ~10 cm, simulating a full discharge in 3D would thus require ten billion grid cells, which exceeds the computational capability of most industrial computing systems. Exploiting azimuthal symmetry to reduce the problem to a 2D system in the R-Z plane lowers these resource requirements considerably. In addition, since the primary physics of the discharge occurs near the cathode (where the magnetic field is strongest and the plasma density is the largest), modeling can be carried out in a truncated simulation domain focused on the cathode/sheath region to further reduce computational costs. Finally, while conventional momentum-conserving PIC schemes suffer from the finite-grid instability when the Debye length is not resolved [22], there has been recent community interest in an explicit energy-conserving PIC method [23] as an alternative approach. Barnes and Chacón [24] have demonstrated that this method (which in this paper will be designated EC-PIC) maintains numerical stability even when the simulation grid does not resolve the electron Debye length, and the method has been successfully utilized in Cartesian geometry by Powis et al.[36] and explored in detail by Adams et al. [25] We have implemented EC-PIC [23] in VSim, and in this work will exercise it in 2D cylindrical coordinates to model dcMS discharges. We will also compare EC-PIC results with the standard momentum-conserving PIC scheme (MC-PIC) [22]. For dcMS discharges where the steady-state plasma density is not known *a priori*, the numerical stability of the EC-PIC method will be shown to be particularly advantageous.

The *explicit* energy-conserving PIC methods we consider here, while related, differ considerably in implementation from the *implicit* energy-conserving PIC methods developed by Refs. [26, 27] Nevertheless, such implicit approaches have also been successfully implemented by other authors to carry out detailed modeling of low-temperature plasma discharges, e.g. in the modeling of planar RF magnetron plasmas [28, 29, 30].

The paper is organized as follows. Section 2 provides a brief explanation of the EC-PIC method. Section 3 describes the simulation setup, boundary and initial conditions, a description of the chemistry included in the model, and the method used to incorporate the background magnetic field. Section 4 describes the general behavior of dsMC systems with examples provided from specific simulations. In Section 5 we benchmark EC-PIC and MC-PIC results and demonstrate improved numerical stability with EC-PIC. In Section 6 we briefly discuss the role of ion-induced secondary electron emission from the cathode on the evolution of the dcMS system. In Section 7 we correlate the sputtered copper flux from the cathode with the ion distribution function impacting the cathode. Section 8 is the conclusion.

**Section 2 – Explanation of MC-PIC and EC-PIC algorithms**

In this section we discuss our implementation of the explicit energy-conserving particle-in-cell method, denoted EC-PIC, in VSim. Lewis [23] first discussed energy conserving PIC in the electrostatic approximation, which is the case we will consider here; Birdsall and Langdon [22] also summarize Lewis' work in their classic textbook. More recently, Barnes and Chacón[32] and Adams et al.[26] have demonstrated the stability of EC-PIC in non-drifting plasmas and provided a stability regime in the case of drifting plasmas. Many problems in low temperature plasma-material interactions (including magnetron sputtering) can be shown to lie in this stable regime and can thus be modeled with EC-PIC; the capacitively coupled plasma discharges modeled by EC-PIC[25, 31] are another example. In this paper, we demonstrate the use of an explicit EC-PIC method in 2D axisymmetric cylindrical simulations for dc magnetron sputtering modeling (dcMS).

The distinguishing difference between EC-PIC and MC-PIC is how fields are interpolated to particle positions. In MC-PIC, the interpolation scheme used to map particles' currents and charges to the simulation grid is identical to the scheme used to interpolate fields to particle positions.[23.] However, in EC-PIC the field-to-particle interpolation method uses a stencil reduced along the direction of the field component by one order, relative to the stencils used for other directions or for the corresponding particle-to-field interpolation. This point is best illustrated by example: in VSim we use linear interpolation (tent functions) to deposit particle charges to the grid. While this linear interpolation scheme is also used for mapping some field components to particle positions, EC-PIC dictates that the Nearest-Grid-Point (NGP) stencil should be used in the field component that is parallel to its grid direction. For example, the interpolation scheme for $E_r(r,z)$ is thus NGP from the $r$-grid to the particle's radial position, and linear from the $z$-grid to its axial position; for the axial field, $E_z(r, z)$, the reverse is true (linear in $r$, NGP in $z$). Higher-order particle stencils (e.g. those of Ref. [32]) can also be used in EC-PIC, but we have chosen to use NGP in this work to allow for easier implementation near the cylindrical axis. We also note that the electric field used in EC-PIC is an "edge-centered" electric field. This means that $E_z(r,z)$ is stored at $[r_m, (z_{n+1} + z_n)/2]$ where $(m,n)$ are nodal grid indices. Likewise, $E_r(r,z)$ is stored at $[(r_{m+1}+r_m)/2, z_n]$. In this context the potential $\varphi$ and the charge density $\rho$ are both stored on the nodal indices so that $\varphi_{m,n} = \varphi(r_m, z_n)$ and $\rho_{m,n} = \rho(r_m, z_n)$.

**Section 3 – Simulation Setup**

We have used a fully kinetic particle-in-cell/Monte-Carlo Collision (PIC-MCC) model in VSim to study the evolution of a dcMS chamber. We confine this study to a 2D R-Z geometry and assume azimuthal symmetry. The model chamber is shown in Figure 1. The simulation domain consists of a region that is $L_Z \times L_R = 3$ cm $\times$ 3 cm. $L_1$ is the radial location of the outer magnet and is fixed in this paper to a value of $L_1 \cong 1.7$ cm. The magnets are configured in an unbalanced type II configuration [2], with a ratio of outer to inner magnetic flux ~ 1.7. The inner magnet has a radius of 6.5 mm while the annular width of the outer magnet is 2 mm. The magnetization for both magnets is identical and is 1.1 Tesla. We have computed the magnetostatic field via a magnetic scalar potential discussed in Jackson [33]. We have assumed Dirichlet boundary conditions. Furthermore, the domain size used to compute the static magnetic field is enlarged compared with the plasma domain to ensure that the boundary conditions do not affect the magnetic field used in the plasma solve. The result of this calculation is shown in Figure 2. The solid black lines indicate the boundaries used in the plasma model.

The electric field boundary conditions are such that the anode is grounded. The anode extends along the z = $Z_{max}$ boundary. The cathode is connected to an external circuit which allows the cathode voltage to float based on (1) the external voltage, and (2) sources and sinks onto the cathode. The black block in Figure 1 that is not part of the cathode is set as a Neumann boundary condition, which allows the voltage to adjust between the cathode and anode. The cylindrical axis is not a physical boundary. However, as part of the field solve, we ensure that Gauss' law is obeyed along this axis.

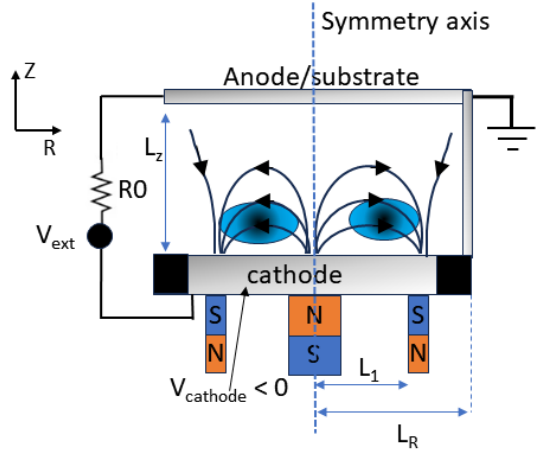

Figure 1: Schematic of the 2D cylindrical geometry used to model magnetron sputtering in VSim.

Absorbing particle boundary conditions are applied on the lower and upper Z-boundaries and the outer radial boundary; the cylindrical axis has a reflective particle boundary condition. Secondary emission from ions impacting the cathode is also included in the model. We have assigned a constant ion-induced secondary electron yield (SEY) for these simulations, with a value of either 0.1 or 0.2 (See Table 1). We also include an electron-induced SEY of 0.1 for electrons impacting the cathode. The electron SEY and ion SEY are independent. Both electron-induced and ion-induced secondary electrons are emitted with zero velocity. However, the electrons are accelerated away from the cathode by the sheath. The model includes neutrals sputtered from the cathode as energetic ions impact its surface. The sputtering yield is based on the model from Ref. [34].

Collisions are modeled using a Monte-Carlo Collision (MCC) scheme [14]. All simulations performed in this study have a background neutral gas pressure of 10 mTorr. We have included four collision processes in the simulations. Although this is a limited set of the full chemistry, these four collisions are sufficient to demonstrate the discharge process and are representative of the fundamental types of collisions that can occur in more complex discharges.

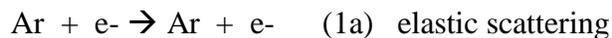

$$Ar + e^- \rightarrow Ar + e^- \quad (1a) \quad \text{elastic scattering}$$

Ar + e- → Ar* + e-     (1b)  excitation

Ar + e- → Ar+ + e- + e-  (1c) ionization from ground state

Ar* + e- → Ar+ + e- + e-  (1d) ionization from metastable state

The excitation energy of collision (1b) is 11.55 eV, which corresponds to one of two metastable excited argon states. The ionization energy in collision (1c) is 15.76 eV. Therefore, the ionization energy in collision (1d) is 4.20 eV, thereby providing discharge mechanism with a lower ionization energy, which is one of the known advantages of using Ar gas. Cross-sections for these 4 collisions are obtained from the QDB database [35]. Furthermore, we use the QDB global model to determine the initial density of Ar+, e-, and Ar*. The input parameters for the QDB model assume a 10 W power source at 10 mTorr. All simulations performed for this paper treat Ar and Ar* as neutral fluids in which only the density evolves. The initial Ar density is $3 \times 10^{20}$/m$^3$ and the initial Ar* density is $2.5 \times 10^{17}$/m$^3$. Over the course of the simulation, the Ar and Ar* densities change very little due to the short time scale of the model. All other species, including sputtered neutrals from the cathode, are treated fully kinetically. A kinetic treatment for Cu is needed to correctly model the Cu flux emitted from the cathode and the deposition of Cu onto the anode.

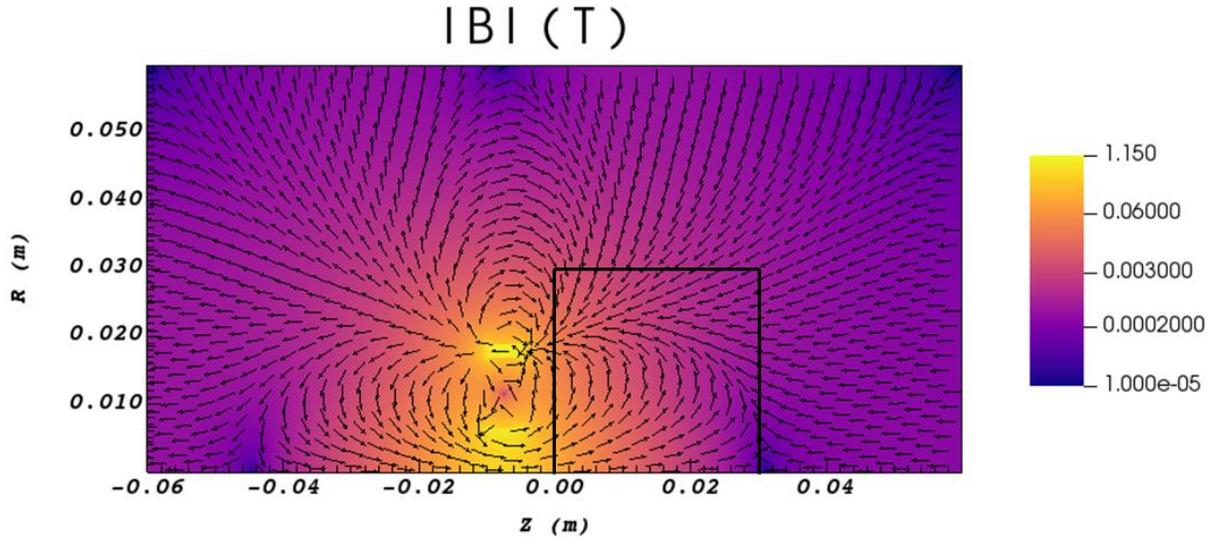

*Figure 2: Magnetic field used to model dcMS in VSim. The simulation domain of the magnetostatic solve extends beyond the simulation domain used for the plasma modeling which are indicated by the solid black lines. The color scale indicates the strength of the magnetic field and the vector field indicates the direction.*

**Section 4 – Magnetron simulation behavior**

The dcMS simulations are initialized with a seed population of uniform density electrons and ions. Most of the electrons are rapidly transported to the anode. However, near the cathode the electrons are confined due to the external magnetic field, and $\vec{E} \times \vec{B}$ drift in the azimuthal direction. Therefore, the lifetime of an individual electron is increased, and it undergoes many more collisions before being absorbed at a boundary compared to an unmagnetized plasma. The discharge grows and is maintained by electron impact ionization and ion-induced secondary electron emission. The latter is effectively diminished by recapture: the magnetic field returns a fraction of emitted electrons to the cathode, a process that is mitigated somewhat by electron-induced secondary electron emission. The sheath that forms near the cathode energizes and repels the secondary electrons into the pre-sheath region (defined below) where each electron can potentially ionize more electrons in electron-neutral ionization collisions.

In dcMS systems, steady state is reached when the production of electrons balances the loss of electrons at the boundaries. One method to hasten convergence to steady state is to supply an external circuit as shown in Figure 1. In this setup, the external voltage is fixed but the cathode voltage is free to float based on the incoming plasma current, external voltage, and resistance. When the total plasma current impinging the cathode is near constant, we say that the simulation has reached steady state. An example of this is shown in Figure 3 where we see that steady state is reached in about 4 µs.

An important physical parameter to measure in these simulations is the electron Debye length. We choose to measure the Debye length in the region where the plasma density is largest, and hence the Debye length is smallest. In the remainder of the paper the smallest Debye length is used as a comparison with the cell size. The sheath typically scales with the electron Debye length. Therefore, more accurate simulations will resolve the Debye length. Nevertheless, we find reasonable agreement even in cases when the Debye length is not well-resolved. Because of the drastic changes in electron density, the Debye length varies considerably in the different regions of the simulation domain.

To illustrate the wide range of plasma densities in the simulation domain, we show in Figure 4 steady state results from EC1 (see Table 1) [note: simulations are labeled EC# or MC#, where EC refers to energy-conserving, MC refers to momentum-conserving and # is a label to identify the simulation]. Figures 4a and 4b show the electron and Ar$^+$ density. Both densities peak at about $3\times10^{16}$/m$^3$ near the cathode sheath ($z < \sim 1.2$ mm), where the ions dominate. The sheath axial location is defined to be the location that the electron plasma density is ½ the ion plasma density. The axial location of the bulk region is where $E_z = d\phi/dz = 0$ ($\phi$ is the potential). The pre-sheath lies between the sheath and bulk regions and is quasineutral. Nevertheless, the pre-sheath has significant electric field and exists roughly between $z = 1.2$ mm and $z = 1$ cm. The pre-sheath is where most of the ionization occurs and is therefore vital to maintaining the plasma. The narrow sheath is seen in the potential shown in Figures 4c and 4d. The three regions (sheath (S), pre-sheath (PS), and bulk) can be seen in Figure 4d, which shows a line plot of the potential through $r \approx 1.2$ cm. In the simulations presented in this paper, the sheath region is where most of the potential drop occurs due the charge separation.

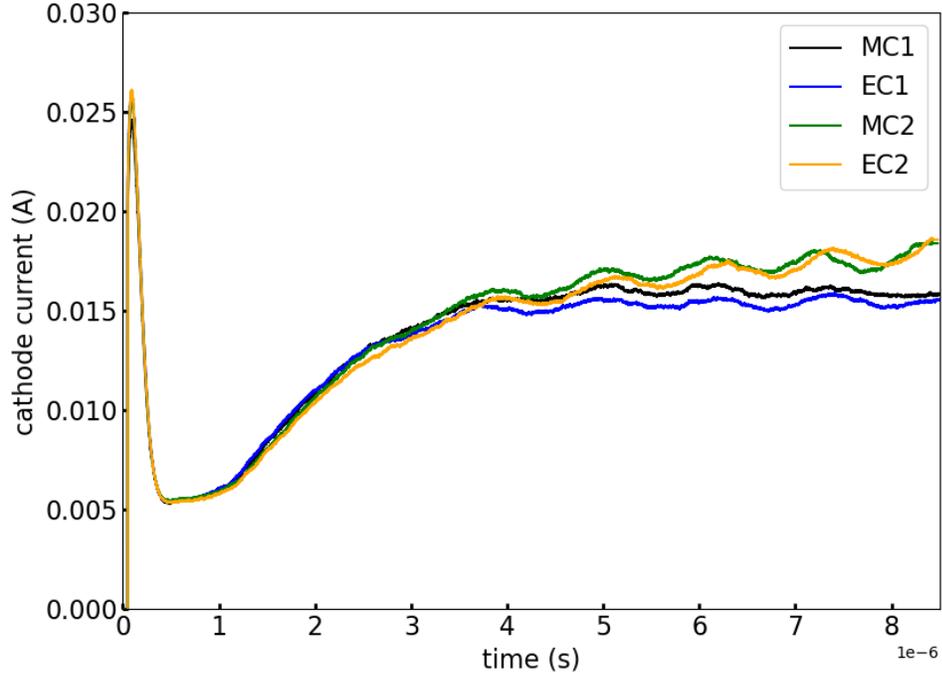

*Figure 3: Cathode current from 4 identical simulations using different PIC schemes and grid sizes. For this plasma regime, the Debye length is well resolved and equivalent results are found in the 4 different runs.*

We have varied several free parameters in carrying out the simulations for this work. Some of these parameters are physics-based (ion-induced SEY, resistance of the external circuit), while other parameters are computational (cell size, computational method). For clarity, we show in Table 1 a summary of the parameters that are varied along with convenient short-names for these parameters.

*Table 1: Table showing the different computational and numerical paramaters that are varied in this paper.*

| | Input Parameters | | | | Steady State Measurements | |
|---|---|---|---|---|---|---|
| Name | Ion-induced SEY | R (Ohms) | Cell size relative to Debye length | method | current (mA) | Potential (V) |
| MC1 | 0.1 | 3000 | 0.5 | MC-PIC | 16 | 250 |
| EC1 | 0.1 | 3000 | 0.5 | EC-PIC | 16 | 250 |
| MC2 | 0.1 | 3000 | 1 | MC-PIC | 18 | 250 |
| EC2 | 0.1 | 3000 | 1 | EC-PIC | 18 | 250 |
| MC3 | 0.2 | 3000 | 1.6 | MC-PIC | 58 | 100 |
| EC3 | 0.2 | 3000 | 1.6 | EC-PIC | 55 | 100 |
| MC4 | 0.2 | 1000 | 5 | MC-PIC | 169 | 90 |
| EC4 | 0.2 | 1000 | 5 | EC-PIC | 165 | 95 |
| MC5 | 0.2 | 1000 | 10 | MC-PIC | -- | -- |
| EC5 | 0.2 | 1000 | 10 | EC-PIC | 166 | 90 |
| EC6 | 0.2 | 1000 | 1.7 | EC-PIC | 167 | 94 |
| EC7 | 0.2 | 1000 | 0.8 | EC-PIC | 167 | 94 |

The SEY for ions impacting the cathode is set to a constant value of either 0.1 or 0.2. The SEY for electrons impacting the cathode is set to 0.1 for all cases. The external circuit parameters are $R0 = 3000\ \Omega$ or $1000\ \Omega$ (resistance) and $V_{ext} = -300$ V (applied DC voltage). In the benchmark simulation, we set $L_z \times L_r = 3$ cm $\times$ 3 cm. The cathode is placed along the $Z = 0$ axis and extends from $R = 0$ to $R = 0.85 * L_r$.

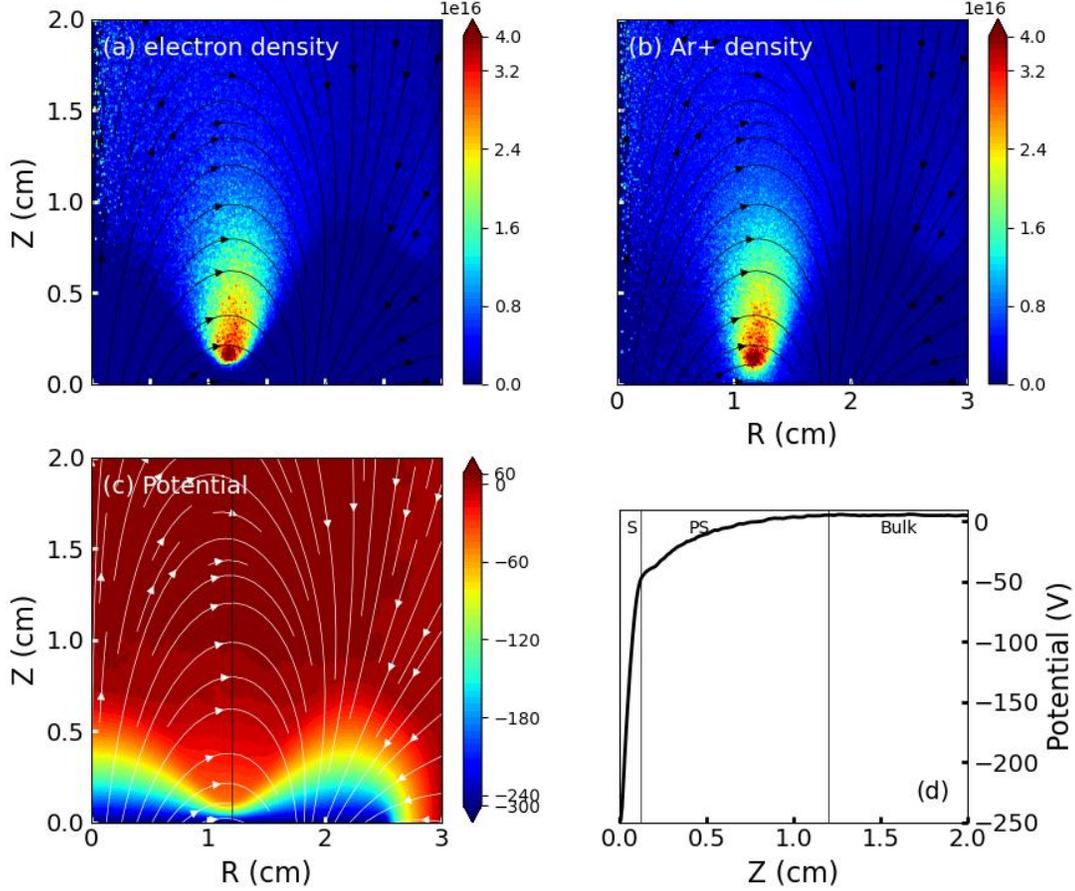

*Figure 4: Electron (a) and Ar+ (b) density show the localization of the plasma discharge region. The charge separation in the sheath that forms near the cathode can be seen by comparing the electron and Ar+ densities. The electric potential is plotted as a function of position in (c) and shown as a radial slice in (d) for Case 1EC. The verical black line in (c) shows the slice taken along the z-direction to produce Figure (d). The potential in the sheath (S), pre-sheath (PS) and bulk plasma regions vary spatially due to the magnetic field confining the plasma. Magnetic field lines are superimposed in (a) – (c).*

## Section 5: The effect of Debye resolution on MC-PIC and EC-PIC

An important parameter is the cell size with respect to the Debye length. In MC-PIC, under-resolving the Debye length leads to numerical instability; in EC-PIC this instability does not exist for stationary plasmas[26]. Thus MC-PIC requires max(dr,dz) < lambda_De, where EC-PIC is stable (though not necessarily accurate) with larger cell sizes. In this section, we explore the accuracy and stability of MC-PIC and EC-PIC as a function of cell size over Debye length. Because EC-PIC does not have the same restrictive criteria as MC-PIC, the former method offers an advantage.

However, because EC-PIC is not as thoroughly tested as MC-PIC and furthermore uses NGP (which is noisier than linear interpolation), it is not known if EC-PIC and MC-PIC agree under identical conditions when $\lambda_{De}$ is well-resolved. Furthermore, it is not clear that EC-PIC will produce the correct answer if $\lambda_{De}$ is poorly resolved. In this section we address both of these issues. We show that EC-PIC and MC-PIC agree when identical simulations are run and when $\lambda_{De}$ is resolved. Furthermore, we show that an under-resolved EC-PIC simulation agrees to within ~15% of a well-resolved simulation (both modeling identical physics). The under-resolved simulation uses ~1/40 the computational resources as the well-resolved simulation, highlighting one advantage of using EC-PIC in magnetron sputtering simulations.

Figure 3 shows the total current impacting the cathode in 4 different cases. MC1 is a momentum conserving simulation with $\max(dr,dz)/\lambda_{De} = 0.5$. EC1 is identical to MC1 except we have used the energy conserving PIC algorithm. MC2 and EC2 set $\max(dr,dz)/\lambda_{De} = 1.0$ but are otherwise identical to their MC1 and EC1 counterparts. For all cases, the particle weight (i.e., the number of physical particles each macroparticle represents) is kept constant. We see in Figure 1 that both EC1 and MC1 reach the same approximate steady state cathode current, indicating the equivalency between EC-PIC and MC-PIC when the Debye length is well resolved. We also see that both MC2 and EC2 show equivalent results. However, the cathode current computed in EC2/MC2 is ~ 20% greater at 8 μs compared with EC1/MC1. Furthermore, in the EC2/MC2 results cathode current is slowly increasing and has yet to reach steady state.

To compare the bulk properties from EC1, MC1, EC2, and MC2, we have computed the average electron temperature after reaching steady state (or nearly reaching steady state, in the case of EC2/MC2) in 3 regions: the sheath, pre-sheath, and bulk. The temperature calculation is also confined to 1.0 cm $< r <$ 1.5 cm, which is the radial bounds of the discharge region. The electron temperature is found by computing the 2$^{nd}$ moment of the particle data, binning the temperature data in the three distinct spatial regions, and averaging the temperature in each region. We find good agreement between the 4 runs, with a slightly higher electron temperature in the sheath region when a larger grid size is used. Note that the sheath is the region with the hottest plasma population. However, because the electron density is small in this region, little ionization occurs there.

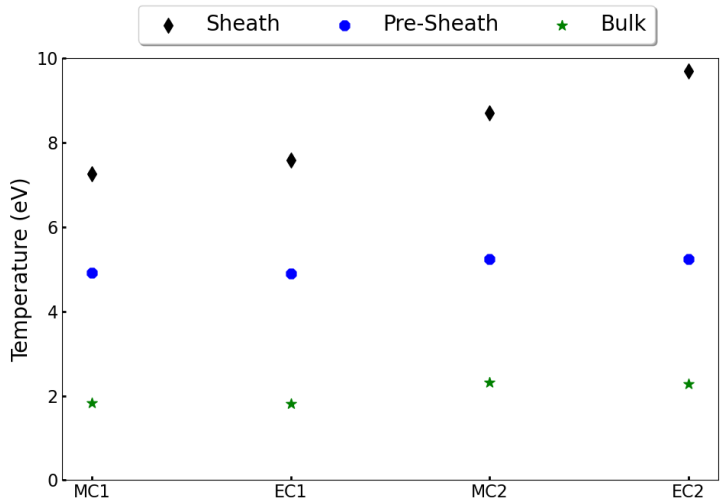

*Figure 5: Electron temperature plotted in 3 different regions from the 4 simulations shown in Figure 2. Good agreement between EC-PIC and MC-PIC, for the spatial resolutions considered, is seen in all 4 simulations.*

To address how well EC-PIC correctly models dcMS when $\lambda_{De}$ is not well-resolved, we have performed a convergence study using EC-PIC with cell-sizes that range from $0.8\lambda_{De}$ to $10\lambda_{De}$. In all cases, the simulation parameters, boundary and initial conditions are identical. The cathode current from four different simulations is shown in Figure 6. In all cases, the steady state cathode current shows good agreement.

We next address how well the sheath and plasma discharge region compare between the simulation with a cell size of $0.8\lambda_{De}$ and $10\lambda_{De}$. Figure 7 shows the electron density from EC7 and EC5. We find that the width of the sheath in EC7 is ~ 0.2 mm. Since the cell size is ~0.019 mm, then there are ~ 10 cells which comprise the sheath in EC7. However, in EC5, the sheath is ~0.5 mm, comprising ~ 3-4 cells. Therefore, with a larger cell size, the sheath expands in order to allow for several cells to resolve the sheath. Furthermore, the electron density is about 20% smaller in EC7 compared with EC5. As we will show in Section 5, the emitted copper flux from the cathode is within 10-20% of the value computed by EC7. The computing resources required to run EC7 are roughly 40x greater than those needed for EC5. We therefore see a trade-off between accurate physics and compute time, but we assert that simulations with fairly coarse spatial resolution can still compute figures-of-merit for dcMS discharges with reasonable accuracy. Achieving optimal discharge conditions in dcMS systems often requires fine-tuning various parameters such as the magnetic field profile and strength. Therefore, by first using a coarse simulation, one can more quickly determine if a given setup will achieve plasma discharge; once a discharge is assured for a given configuration, the model can be further refined.

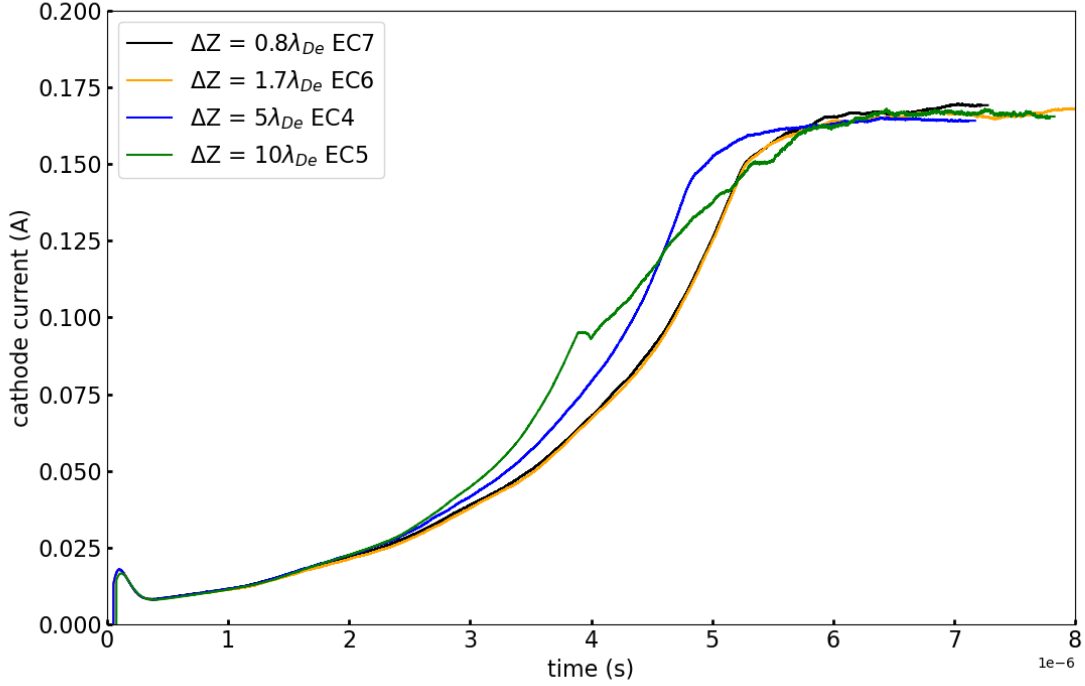

*Figure 6: Cathode current from 4 different EC-PIC simulations of various resolutions. In each simulation, the initial conditions, boundary conditions, and process parameters are identical. We see good agreement in the steady-state current in all 4 cases.*

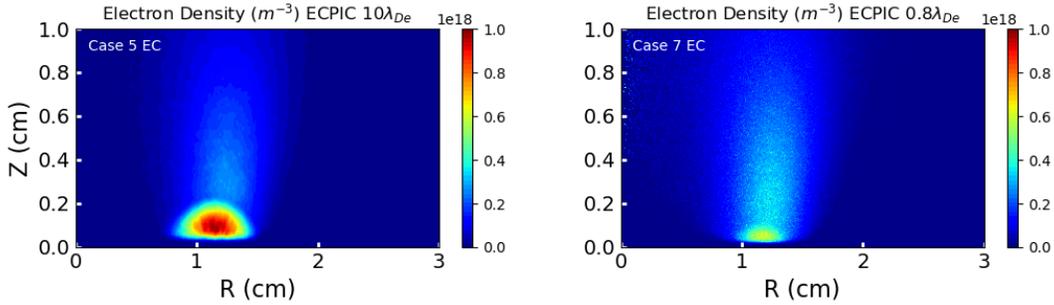

*Figure 7: Electron density from Case5 EC (left) and Case 7 EC (right).*

Finally, we compare EC-PIC and MC-PIC in cases that the Debye length is under-resolved. MC4 and MC5 are identical to EC4 and EC5 except that the former 2 cases use MC-PIC and the latter two cases use EC-PIC. Figure 8 shows the cathode current from the 4 simulations. In the runs with larger grids (EC5 and MC5), we initially ran the simulation with twice as large a time step relative to the small grid size runs. However, we found that this leads to numerical instability once the density exceeds ~5x$10^{17}$ m$^{-3}$, which occurs at around 4 μs. At the same time, we also found that peak particles-per-cell counts exceed 1000. To reduce the computational cost, we employed a particle re-weighting scheme in which we removed 1/3 of the particles and increased the particle weight by 3. We also reduced the time step to be equivalent to the small cell size run. After making these adjustments, we found that EC5 reached steady state with a current nearly equal to that of the small ΔZ run. We also observe that EC-PIC simulations are more stable than the MC-PIC runs.

The large ΔZ MC-PIC run (MC5) is numerically unstable since grid heating (arising from the under-resolved electron Debye length) leads to excessive electron temperatures.

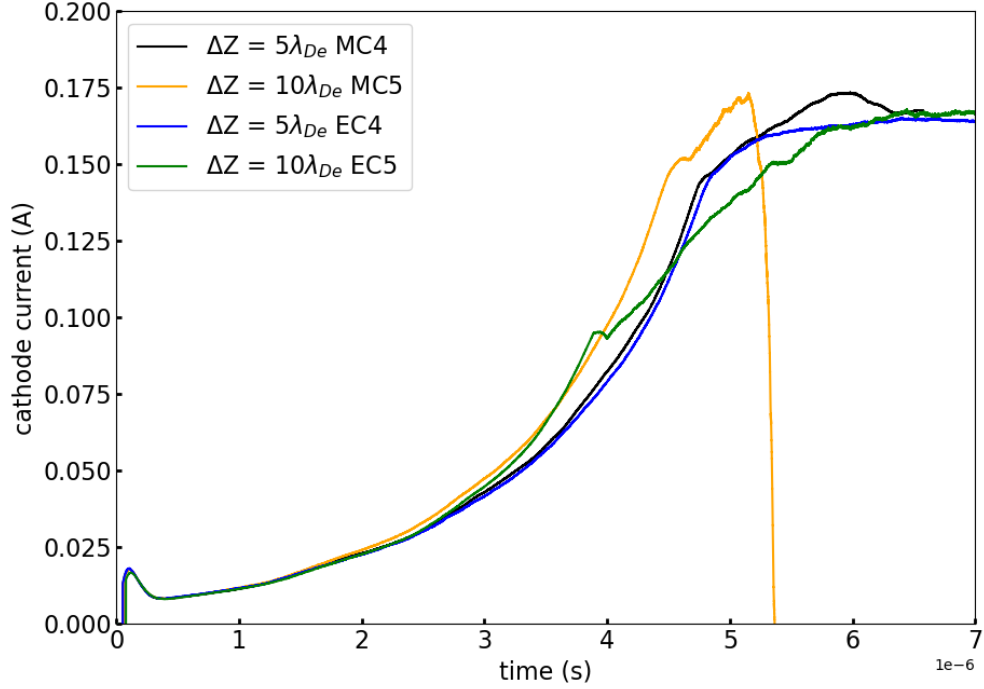

*Figure 8: Cathode current from four simulations with SEY = 0.2 and R0 = 1000 Ω. Initial and boundary conditions are the same in all four cases. The black and blue curves are from MC-PIC and EC-PIC with ΔZ = ΔR = 0.076 mm. The green and orange curves are from MC-PIC and EC-PIC simulations with the grid size doubled.*

**Section 6 - Effect of varying SEY and external resistance**
*Varying SEY*
 In this section, we explore how changing the ion-induced SEY impacts the steady-state current and plasma density. We examine ion-induced SEY values of 0.1 and 0.2 and leave all other parameters unchanged; 0.2 is the upper limit of typical SEY for metallic surfaces [2]. Physically, this change is equivalent to choosing a different cathode material, or modeling surfaces that contain different levels of surface contaminants. For the remainder of the paper, we will show results using only the EC-PIC method. We ran simulations using the following circuit parameters: R = 3000 Ω and $V_{ext}$ = -300 V. Figure 9 shows a comparison of cathode current for the SEY = 0.1 and SEY = 0.2 cases. Larger SEY naturally leads to a larger cathode current since each ion impacting the cathode has higher probability of emitting an electron. Furthermore, the additional electrons cause an increase in the ionization rate of the source gas, leading to an increase in plasma density. We show in Figure 10 the steady-state electron density in the SEY = 0.1 (EC1) and SEY = 0.2 runs (EC3).

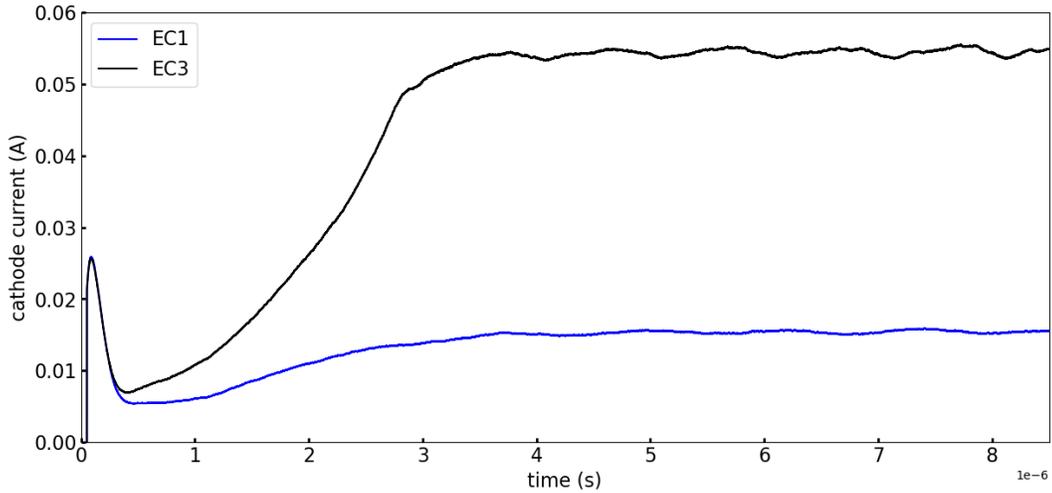

*Figure 9: Comparison of cathode current from 2 different values of SEY.*

There are notable differences between the 0.1 SEY and 0.2 SEY runs. In addition to the larger current corresponding to a larger plasma density, the sheath is much smaller (as shown in Figure 10) in the larger SEY runs. Inspection of the electron density shown in Figure 10 shows that the sheath is ~ 1 mm thick in the 0.1 SEY run and ~0.25 mm in the 0.2 SEY run. We also find that the steady-state cathode voltage is smaller for higher secondary yields; in the high-yield (0.2 SEY) case we obtain a cathode voltage ~ -100 V, while in the low-yield (0.1 SEY) case this voltage is ~ -250 V.

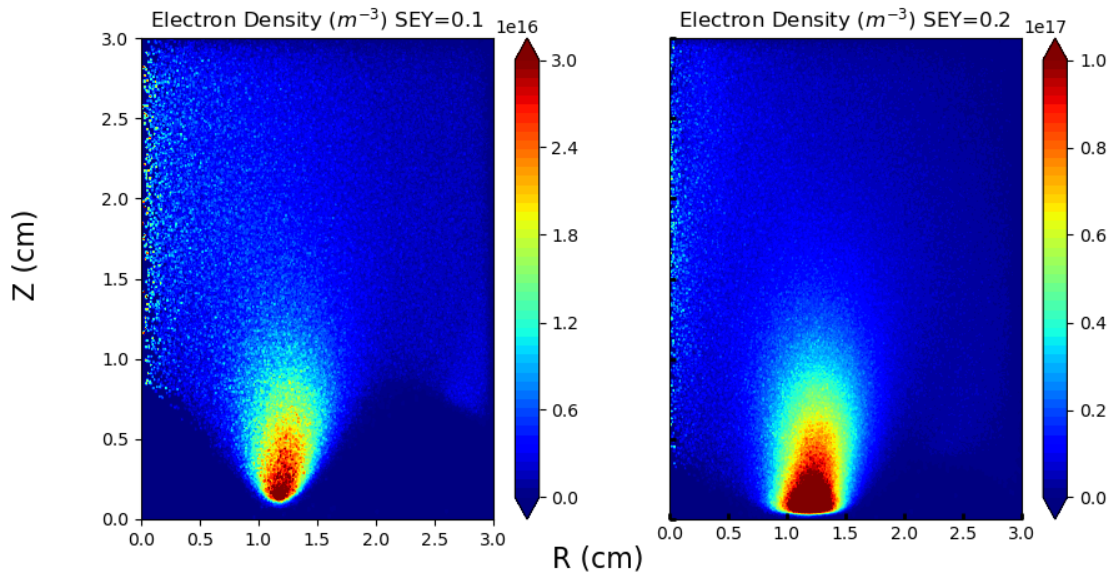

*Figure 10: Comparison of electron density from two different runs with different values of the SEY. Both plots are from EC-PIC simulations are represent Case1 EC (left plot) and Case3 (EC) (right plot).*

*Role of External Circuit*

We next modify the external circuit by decreasing the external resistance from 3000 Ω to 1000 Ω. By keeping the external voltage at -300 V, we can determine how this resistance impacts the steady-state current and plasma density. The results of this study can be seen in Table 1. For example, comparing EC3 with EC6 (both have about the same cell size and the ion-induced SEY is 0.2 in both cases), we see that a smaller external resistance leads to a larger steady state current and a correspondingly larger plasma density. The external circuit probes different points on the IV curve. By varying the external resistance, we found that our simulated device had a slightly negative differential resistance. Comparison of Figure 5 and Figure 7 shows an order of magnitude increase in density when the external resistance changes from 3000 Ω to 1000 Ω. Therefore, modeling larger currents can be a challenge for kinetic codes due to the need to resolve the electron Debye length. EC-PIC provides one solution to this issue by allowing for larger cell sizes relative to the electron Debye length.

**Section 7 – Comparison of Ion Distribution on cathode and Cu Flux emitted from cathode**

A quantity of interest in most applications is the deposition rate of the sputtered cathode material onto the substrate. Methods for computing the deposition rate will be discussed in more detail in a companion paper; for this work, since the deposition rate depends on the $Ar^+$ ions impacting the cathode, it will suffice to examine argon ion energy distributions (IEDs) at the cathode. Because of the increased stability seen in the EC-PIC runs, we will compare IEDs from EC1, EC3, and EC4. We will also compare the copper flux off the cathode for these three runs. There is significant variation in current between these three cases, and because the external voltage is fixed, the increased current leads to a decrease in cathode voltage. The SS voltages/currents in EC1, EC3, and EC4 (respectively) are 250 V/15 mA, 100 V/55 mA, and 95 V/165 mA. Comparing the last two simulations will demonstrate the role of the current (for a given voltage) on the emitted Cu flux.

Figure 11 shows the ion energy distribution for the three cases. The data shown in Figure 11 are computed by binning the data along the cathode between 1.0 cm and 1.5 cm. As expected, we observe larger incident energies for larger sheath potentials, but the plots all have common features. The high-energy peak is a measure of the total voltage drop, while the low-energy peak is a measure of the sheath voltage. The spread of energy between the high and low peaks occurs because ions are created between the sheath and anode and therefore experience a range of potential drops. We see in Figure 12 an increase of ions in the sheath in EC3 and EC4 compared with EC1. This is likely due to the greater source of ionizing electrons which are emitted from the sheath in EC3 and EC4 due to the increased ion-induced SEY.

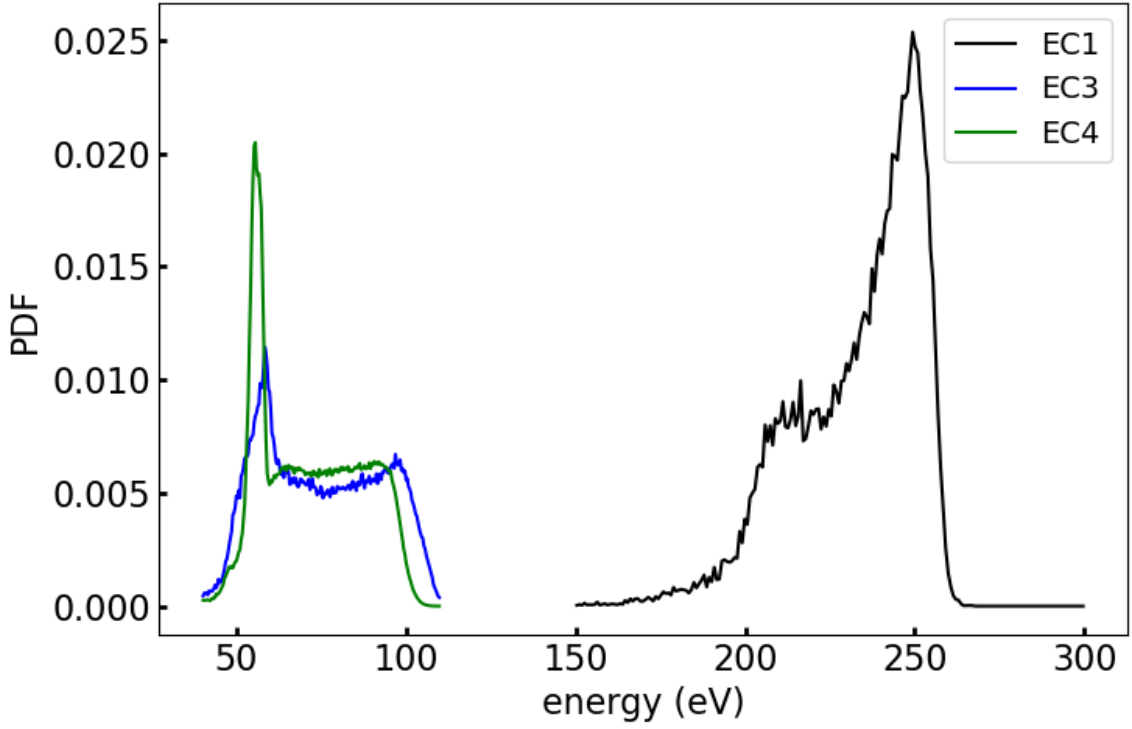

*Figure 11: Ar+ energy distribution function at the cathode for three cases with varied currents and voltages.*

The next question we address is how the sputtered Cu flux, generated as ions sampled from these IEDs strike the cathode, depends on sheath voltage and cathode current. Figure 12 shows that higher currents increase the copper flux. EC3 and EC4 have nearly the same cathode voltage. However, the current in EC4 is ~3 times larger, with a corresponding 3x increase in the Cu flux. Finally, we provide one more result highlighting the advantage of the coarse cell EC-PIC model. We show in Figure 13 the emitted copper flux from the 4 runs shown in Figure 8, i.e. the 4 runs used in the convergence study. We observe that the steady state copper flux emitted from the cathode agrees within ~15% between the $10\lambda_{De}$ and $0.8\lambda_{De}$ simulations. It therefore appears plausible that EC-PIC methods can greatly reduce the computational resources needed to make reasonable estimates of steady-state dcMS discharge conditions.

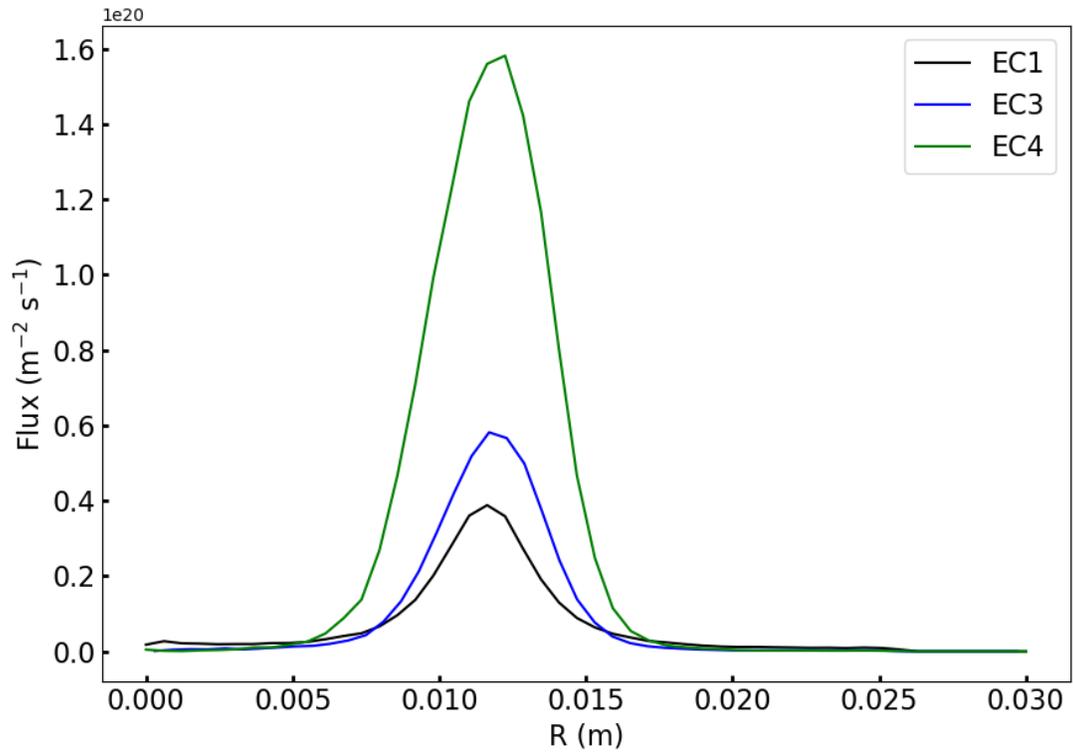

*Figure 12: Cu flux emitted from the cathode from the three different runs shown in Figure 10.*

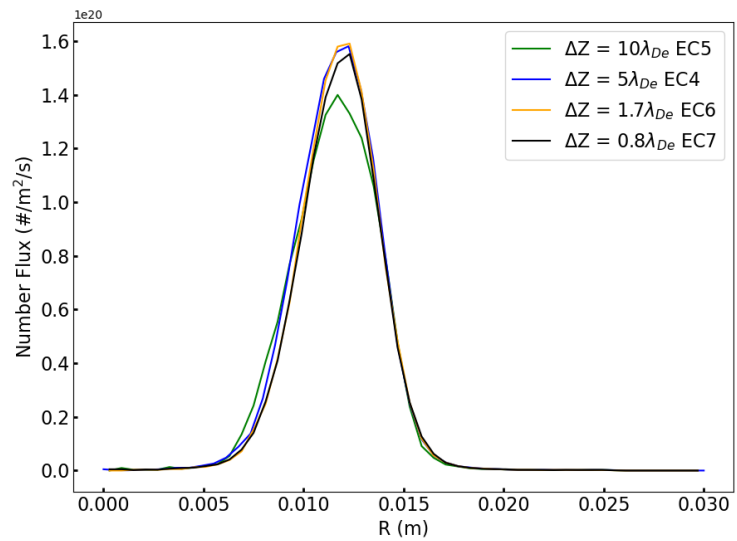

*Figure 13: Copper flux emitted from the cathode from 4 different runs. Runs differ only in the cell size, with all other initial conditions, boundary conditions, and plasma conditions identical.*

**Conclusion**

This paper has discussed results from a cylindrically symmetric dc magnetron sputtering simulation in which the working gas is argon and the cathode is modeled as copper. We have demonstrated a new implementation of an energy-conserving PIC algorithm in cylindrical geometry. While EC-PIC has been around for ~50 years, it has not gained widespread usage. However, there has been recent discussion of its use in Cartesian geometry. We have presented an implementation of EC-PIC in cylindrical geometry and have shown that it exhibits improved numerical stability compared with the standard MC-PIC method. We have demonstrated that in cases where the electron Debye length is not well resolved, MC-PIC is not suitable due to grid heating, as seen in Figure 8. Furthermore, we have shown agreement within ~15% of the emitted copper flux between under-resolved and well-resolved simulations when EC-PIC is used. In addition, we also realize a factor of ~40 reduction in computational resources between under-resolved and well-resolved simulations. These results suggest a potential workflow for scoping out dcMS discharge configurations – process parameters that lead to discharge can be quickly explored using coarsely-resolved EC-PIC simulations, after which the model can be refined as allowed by available computing resources.

In addition to exploring differences between EC-PIC and MC-PIC, we also explored the role of ion-induced SEY on the plasma evolution and ion energy distribution for steady-state dcMS discharges. We show that increasing the SEY (without changing the external circuit) leads to a greater SS current while also decreasing the cathode voltage. We also demonstrate that the ensuing sputtering flux of neutral copper atoms depends primarily on the incident $Ar^+$ flux and is less dependent on the energy. This is demonstrated in Table 1 by comparing Case 3 and Case 4; while both cases have similar cathode voltages (~ 100 V), EC4 has 3x the SS current as EC3 and Figure 13 consequently shows a sputtered flux that is 3x larger than EC3. Finally, we show that reducing the external resistance increases the SS current and reduces the SS voltage, which leads to a larger SS discharge density.